\documentclass[prx,twocolumn,floatfix,superscriptaddress,amsmath,aps,longbibliography]{revtex4-2}
\pdfoutput=1
\usepackage{dcolumn,graphicx,color,booktabs,microtype,afterpage,bm}
\makeatletter\renewcommand{\fnum@figure}[1]{\figurename~\thefigure.}\makeatother
\makeatletter\renewcommand{\fnum@table}[1]{\tablename~\thetable.}\makeatother
\usepackage[colorlinks,plainpages=false,linkcolor=blue,urlcolor=blue,citecolor=blue,pdfpagemode=UseNone,pdfstartview=FitBH]{hyperref}

\newcommand{\NFS}{NaFeSi$_2$O$_6$}

\begin{document}

\title{Spin dynamics in natural multiferroic pyroxene NaFeSi$_2$O$_6$}

\author{Oleksandr Prokhnenko}
\affiliation{Helmholtz-Zentrum Berlin f\"ur Materialien und Energie, Hahn-Meitner-Platz 1,
Berlin D-14109, Germany}
\author{Stanislav E.~Nikitin}
\affiliation{PSI Center for Neutron and Muon Sciences, Paul Scherrer Institut, CH-5232 Villigen-PSI, Switzerland}
\author{Koji Kaneko}
\affiliation{Materials Sciences Research Center, Japan Atomic Energy Agency, Tokai, Ibaraki 319-1195, Japan}
\affiliation{Advanced Science Research Center, Japan Atomic Energy Agency, Tokai, Ibaraki 319-1195, Japan}
\author{Chihiro Tabata}
\affiliation{Materials Sciences Research Center, Japan Atomic Energy Agency, Tokai, Ibaraki 319-1195, Japan}
\affiliation{Advanced Science Research Center, Japan Atomic Energy Agency, Tokai, Ibaraki 319-1195, Japan}
\author{Yusuke Hirose}
\affiliation{Materials Sciences Research Center, Japan Atomic Energy Agency, Tokai, Ibaraki 319-1195, Japan}
\affiliation{Advanced Science Research Center, Japan Atomic Energy Agency, Tokai, Ibaraki 319-1195, Japan}
\author{Yoshifumi Tokiwa}
\affiliation{Advanced Science Research Center, Japan Atomic Energy Agency, Tokai, Ibaraki 319-1195, Japan}
\author{Yoshinori Haga}
\affiliation{Advanced Science Research Center, Japan Atomic Energy Agency, Tokai, Ibaraki 319-1195, Japan}
\author{Masaki Fujita}
\affiliation{Institute for Materials Research, Tohoku University, Sendai, 980-8577, Japan}
\author{Hiroyuki Nojiri}
\affiliation{Institute for Materials Research, Tohoku University, Sendai, 980-8577, Japan}
\author{Lawrence M.~Anovitz}
\affiliation{Chemical Sciences Division, Oak Ridge National Laboratory, Oak Ridge, TN 37831, USA}
\author{Andrey~Podlesnyak}
%\thanks{Corresponding author: podlesnyakaa@ornl.gov}
\affiliation{Neutron Scattering Division, Oak Ridge National Laboratory, Oak Ridge, Tennessee 37831, USA}

\begin{abstract}
Spin dynamics in the natural mineral aegirine, NaFeSi$_2$O$_6$, a member of the pyroxene family, was studied by elastic and inelastic neutron scattering.
Magnetization and specific heat measurements as well as single-crystal neutron diffraction maps, taken in the temperature range 2 -- 20~K, confirm two successive magnetic transitions at 8.8 and 5.8~K, consistent with previous studies.
The observed spin-wave excitations emerge from the incommensurate magnetic Bragg peaks corresponding to the propagation vector $k_{\rm ICM} = (0, 0.77, 0)$, and extend up to energies of about 1.5~meV.
In the low-temperature helical phase, the spin dynamics of the Fe$^{3+}$ ions is well described by a simple linear spin-wave model.
The observed excitations can be modeled using a spin Hamiltonian that includes three primary exchange interactions - intrachain coupling $J=0.142(2)$~meV, interchain couplings $J_1=0.083(1)$~meV and $J_2=0.186(1)$~meV - and an easy-plane anisotropy $D=0.020(6)$~meV.
Our results show that no single exchange interaction dominates the spin dynamics. 
The similar strengths of the intrachain and interchain couplings point to the fact that the magnetic interactions in aegirine are three-dimensional rather than confined along one direction.
As a result, the system cannot be considered quasi-one-dimensional, as previously suggested, and calls for further investigations. 
\end{abstract}

\maketitle{}

\section{Introduction}
\label{Intro}

Antiferromagnets with competing interactions remain an important topic in condensed matter physics. Due to the interplay between different exchange paths, these systems exhibit a variety of unusual physical properties, including magnetization plateaus~\cite{takigawa2010magnetization}, unconventional magnetic excitations~\cite{li2017detecting, kamiya2018nature, xie2023complete}, quantum criticality~\cite{vojta2018frustration, fan2024field} and multiferroicity~\cite{kimura2003magnetic, arkenbout2006ferroelectricity, choi2008ferroelectricity} to mention a few.
Thus, these materials attract significant interest, not only for their potential practical applications but also for the fundamental insights they offer.

The pyroxene family, a large group of natural minerals and inorganic compounds with the general formula $ATX_2$O$_6$ ($A$ = Sr, Li, Na, or Ca; $X$ = Ge or Si; and $T$ represents a magnetic transition metal), is an example of such systems.
Most pyroxenes exhibit rich low-temperature magnetic behavior, partly due to the geometry of the magnetic lattice, where chains of 3d transition-metal ions are coupled in a way that forms quasi-triangular motifs within certain crystallographic planes.
Several pyroxene compounds have recently been found to exhibit magnetoelectric (ME) behavior.
Examples include linear ME in materials like LiCrSi$_2$O$_6$~\cite{Nenert,Janson}, LiFeSi$_2$O$_6$~\cite{Jodlauk_2007,Baum2013}, CaMnGe$_2$O$_6$~\cite{Ding}, as well as spin-driven multiferroicity in NaFeSi$_2$O$_6$~\cite{Streltsov_2010,Redhammer2011,Baum2015}, NaFeGe$_2$O$_6$~\cite{Redhammer2011,Biesenkamp2021}, and SrMnGe$_2$O$_6$~\cite{Ding_2016}.
Recently, the Co-based pyroxene SrCoGe$_2$O$_6$ has also been suggested as a promising candidate for the realization of the Kitaev model due to the strong spin-orbit coupling of Co$^{2+}$ ions~\cite{Maksimov}.

\begin{figure}[bt]
\center{\includegraphics[width=1.0\linewidth]{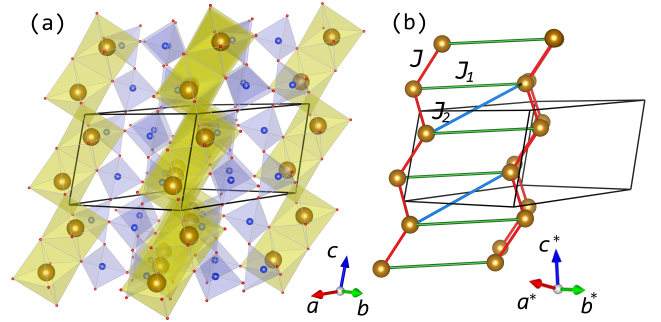}}
  \caption{~(a) Crystal structure of \NFS\ drawn by \textsc{Vesta}~\cite{vesta}. Edge-sharing FeO$_6$ octahedra and corner-sharing SiO$_4$ tetrahedra are shown in yellow and blue, respectively. Sodium atoms are not shown for clarity. (b) The three primary exchange interactions -- intrachain $J$ and interchain $J_1$, $J_2$ -- are indicated.  Axes in direct and reciprocal space are shown on the left and right compasses, respectively.
  }
  \label{fig_structure}
\end{figure}

The natural single crystals of aegirine \NFS used in this study, although they show a slight deviation from the ideal composition (which we discuss in Sec.~\ref{sec::characterization}), were found to be multiferroic~\cite{Jodlauk_2007}.
\NFS\ crystallizes in the monoclinic space group $C2/c$ with cell dimensions $a=9.68$~\AA; $b=8.83$~\AA; $c=5.30$~\AA; $\beta =107.3^{\circ}$~\cite{Clark1969,Ballet1989}.
As shown in Fig.~\ref{fig_structure}(a), the structure has iron in zigzag chains of edge-sharing (FeO$_6$) octahedra and chains of corner-sharing (SiO$_4$) tetrahedra that run along the $c$-direction.
The arrangement of the magnetic chains forms a triangular magnetic lattice within each (110) plane, resulting in magnetic frustration.
Spin sublattice can be described by chains with intrachain coupling $J$, linked by two interchain couplings $J_1$ and $J_2$ [Fig.~\ref{fig_structure}(b)]~\cite{Jodlauk_2007}.
These three primary couplings create a triangular topology and a spin-frustrated network of iron spin chains.
A detailed analysis of the variety of magnetic phase diagrams in pyroxenes as a function of the intrachain exchange interaction $J$, was carried out in Ref.~\cite{Streltsov2008} using both perturbation theory and band structure calculations.
Although some authors have proposed the importance of frustrated interchain couplings, $J_1$ and $J_2$, based on geometric considerations~\cite{Streltsov2008,Nenert2010}, this idea has not been thoroughly examined. 
A detailed microscopic investigation of the magnetic models is still lacking.

On cooling, \NFS\ exhibits two magnetic transitions in incommensurate (ICM) magnetic order. Between 9 and 6~K it has a transverse spin-density wave with moments pointing near the $c$-direction. Below 6~K, magnetic order
becomes helical and the spins rotate in the $ac$-plane~\cite{Baum2015}. 
The Curie-Weiss temperature deduced from magnetization measurements, $T_{\rm CW} = -39$~K~\cite{baum1988structural}, significantly exceeds the ordering temperature yielding the frustration ratio $f = |T_{\rm CW}/T_{\rm N}| = 4.3$. This indicates the presence of magnetic fluctuations well above the ordering temperature, which is typically associated with low-dimensionality or frustrations.
Moreover, natural \NFS\ is a rare example of a multiferroic material whose polarization, $\mathbf{P}$, cannot be described by the conventional spin current mechanism, $\mathbf{P} \propto \mathbf{r}_{ij} \times (\mathbf{S}_i\times \mathbf{S}_j)$, where $\mathbf{r}_{ij}$ is the connecting vector between two neighboring spins $\mathbf{S}_i$ and $\mathbf{S}_j$ \cite{SpinCurrent}.
Instead, the plane of the magnetic helix and the direction of the electric polarization are orthogonal, $\mathbf{P} \propto \mathbf{S}_i\times \mathbf{S}_j$~\cite{Baum2015}.

In this paper, we use elastic and inelastic neutron scattering (INS) to explore the low-energy magnetic excitations in the ICM helical phase of \NFS.
By applying linear spin-wave theory (LSWT), we were able to interpret the INS data and describe the observed excitation spectrum. 
This approach also allowed us to build an effective spin Hamiltonian and determine the key exchange interactions, offering deeper insight into the competing interactions that give rise to the complex magnetic order of the material.
Although earlier studies suggested that \NFS\ exhibits quasi-one-dimensional magnetism due to its chain-like crystal structure, our results reveal a markedly different picture.
We find that the interchain interaction is not only substantial, but even exceeds the intrachain coupling. 
This unexpected hierarchy of exchange interactions reveals the three-dimensional nature of magnetism in aegirine and underscores the importance of a full microscopic analysis to understand its complex magnetic behavior.

\section{Experimental methods}
\label{method}

The natural single crystal of aegirine,  \NFS, used for the experiments described in this paper was obtained commercially. 
It was found at Mount Malosa, Zomba, Southern Region, Malawi.
The elemental composition of the crystal was characterized using an electron-probe microanalyzer (JXA iHP200F, JEOL Ltd.). Characteristic x-rays emitted from the sample, averaged over a circular area with a radius of 0.01 mm, were analyzed by four wavelength-dispersive spectrometer crystals: LDE2 (Layered Diffracting Element, optimized for low-energy x-rays), TAP (thallium acid phthalate), PET (pentaerythritol), and LiF (lithium fluoride).

Magnetization measurements were performed by sample extraction method using a SQUID magnetometer (MPMS, Quantum Design) under a magnetic field of up to 7 T between 2 and 300 K. 
Specific heat measurements were carried out by means of the heat relaxation technique with the Physical Property Measurement System (PPMS, Quantum Design).

The neutron diffraction experiments were conducted using the triple-axis spectrometer TAS-2 at the JRR-3 research reactor of the Japan Atomic Energy Agency in Tokai, Japan. The spectrometer was operated in elastic mode with incoming and outgoing energy of $E = 14.7$~meV. Higher-order contamination was suppressed using a pyrolytic graphite filter.
The sample was cooled down by a dry-type $^4$He top-load cryostat\cite{CLVTC2024}.

INS experiments were performed on the time-of-flight (TOF) Cold Neutron Chopper Spectrometer (CNCS)~\cite{CNCS1,CNCS2} at the Spallation Neutron Source, Oak Ridge National Laboratory.
For the first CNCS experiment, the 3.4 gram crystal was mounted on an aluminum sample holder and aligned in the $(h~k~0)$ scattering plane.
For the second, the sample was aligned in the $(0~k~l)$ scattering plane.
All measurements were performed at a temperature of 1.7~K using a standard orange cryostat.
To analyze the low-energy spin dynamics in detail, the INS data were collected with incident energy $E_{\rm i} = 3.32$~ meV, which yields an energy resolution of 0.11~meV at the elastic line. Using the rotating single-crystal method, all TOF data sets were combined to construct a four-dimensional scattering intensity function, $I(\mathbf{Q},\hbar\omega)$, where $\mathbf{Q}$ represents the momentum transfer and $\hbar\omega$ denotes the energy transfer.
The software packages \textsc{Horace}~\cite{Horace}, and \textsc{MantidPlot}~\cite{Mantid} were used for data reduction and analysis, while the SpinW software was employed for LSWT calculations \cite{SpinW}.
The Fe$^{3+}$ magnetic form factor and the instrumental resolution was included in the calculations.

Preliminary inelastic sample characterization was done using the TASP triple-axis spectrometer at SINQ (Paul Scherrer Institute) operated with $E_{\rm f} = 3.5$~meV. A cold Be filter was used to suppress higher-order harmonics.

\section{Results and analysis}

\subsection{Sample characterization}
\label{sec::characterization}

\begin{figure}[tb]
\center{\includegraphics[width=1.0\linewidth]{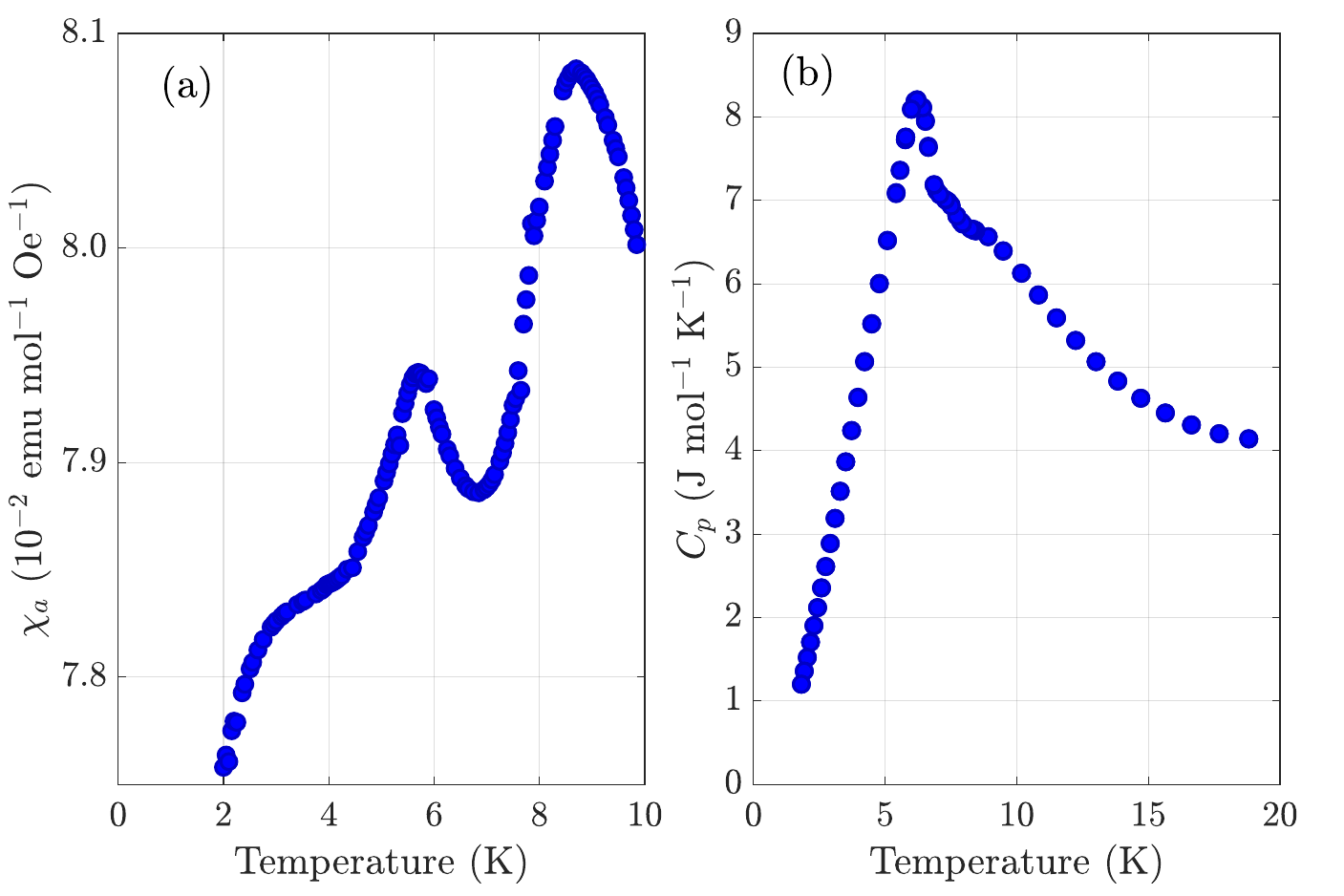}}
  \caption{~(a) Temperature dependence of the static magnetic susceptibility $\chi_a=M/B$ of \NFS\ in a magnetic field $B=0.1$~T along the $a$ direction, reveling two magnetic transitions around 6 and 9~K. (b) Low-temperature heat capacity $C_p$($T$) of \NFS recorded in zero magnetic field.
  }
  \label{fig_hc}
\end{figure}

The magnetic properties of natural aegirine have been investigated by several research groups, and our bulk measurements are consistent with the findings reported in Refs.~\cite{Jodlauk_2007,Baker2010,Baum2015}.
Figure~\ref{fig_hc}(a) shows the temperature dependence of the static magnetic susceptibility $\chi_a$($T$) of \NFS\ measured in a magnetic field of 0.1~T applied parallel to the $a$-axis.
The peaks in the susceptibility curve indicate the presence of two successive magnetic transitions at temperatures  5.8(1)~K and 8.8(1)~K. These transitions are further confirmed by the heat capacity data $C_p$($T$) displayed in Fig.~\ref{fig_hc}(b), where anomalies are observed at the same temperatures. 

Note that the chemical composition of the natural single crystal used in this study deviates slightly from the stoichiometric one and from the compositions reported by other research teams~\cite{Jodlauk_2007,Baum2015}.
According to our composition analysis, the determined sample stoichiometry is Na(Fe$_{0.96}$Ti$_{0.03}$Mg$_{0.01}$)(Si$_{1.99}$Al$_{0.01}$)O$_6$), suggesting that the main magnetic properties remain consistent across the family of natural aegirine, despite these slight variations in elemental composition. 
However, we note that the synthetic \NFS~behaves differently. 
In contrast to the ICM order observed in natural samples, the synthetic crystal showed a coexistence of an ICM order with a commensurate magnetic structure and lacked ferroelectricity~\cite{Baker2010,Baum2015}.
These differences point to fine changes in the magnetic structure that have a significant impact on the ferroelectric order. Although further investigations in this direction would be helpful, they remain beyond the scope of the current work.

%\subsection{Neutron Diffraction}
%\label{sec::ENS}

Before turning to the INS results, we summarize the sample characterization using elastic neutron scattering. 
Our results show very good agreement with those reported in Ref.~\cite{Baum2015}. 
Single-crystal neutron diffraction was performed in the ($h~k~0$) scattering plane at different temperatures below and above the magnetic ordering temperature, $T_{\mathrm{N}}$ = 8.8~K. 
At a base temperature of $T\!=\!2.5$~K, strong superlattice magnetic peaks are detected %\blue{
around the allowed nuclear peaks, indicating a well-defined magnetic phase. 
These magnetic peaks can be indexed with a single ICM propagation vector $k_{\rm ICM}\!=\!(0~0.77~0)$.
%, in agreement with previous reports~\cite{Baum2015}.

\begin{figure}[tb]
\center{\includegraphics[width=1.0\linewidth]{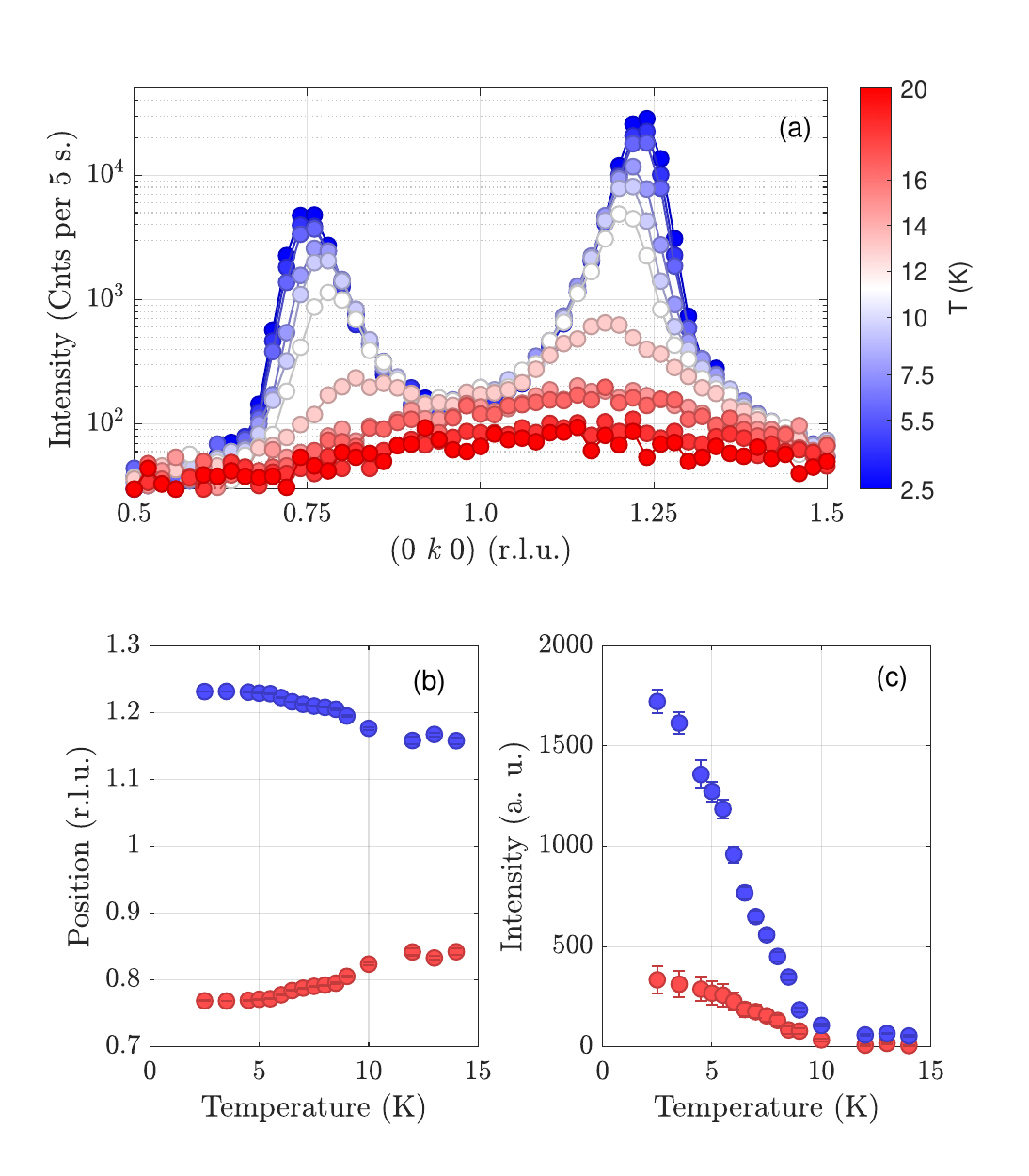}}
  \caption{~(a) Elastic scans along $(0~k~0)$ direction at several representative temperatures across the magnetic transitions. (b) Temperature dependence of the ($0~k\!\pm\!k_{\rm ICM}~0$) peak positions and (c) their intensities.
  Magnetic intensity becomes resolution-limited below $\sim$6~K. The weak Q-dependent modulation may reflect magnetic fluctuations above $T_{\mathrm{N}}$.
   }
  \label{fig_elastic}
\end{figure}

Figure~\ref{fig_elastic}(a) visualizes a set of temperature-dependent scans along the ($0~k~0$) direction, representing the evolution of the magnetic Bragg peaks with temperature. 
Magnetic peaks become clearly visible near the N\'{e}el temperature ($T_{\mathrm{N}}=8.8$~K). 
The intensities increase with cooling down to the base temperature with a small step-like anomaly across the second magnetic transition at 5.8~K [Fig.~\ref{fig_elastic}(c)]. 
This has been interpreted as the emergence of a helical spin-spiral structure from a spin-density wave~\cite{Baum2015}.

With decreasing temperature the peak position continuously shifts [Fig.~\ref{fig_elastic}(b)], reflecting a change in the propagation vector $k_{\rm ICM}$. Below roughly 5~K however there is no more temperature-driven modulation of the magnetic ordering wave vector, indicating a lock-in transition often observed in multiferroic systems; see e.g. Refs. \cite{Biesenkamp2021, tbmno3, Radaelli_2008}.

In addition to the sharp incommensurate magnetic Bragg peaks, a weak diffuse scattering signal is observed along the $k$-direction between neighboring peaks, as seen in Figs.~\ref{fig_elastic}(a) and \ref{fig_elastic_tof}(a). 
This diffuse scattering likely reflects short-range magnetic correlations or slight inhomogeneities in the incommensurate modulation, possibly arising from local compositional variations or strain in the natural crystal. 
Given its low intensity, the signal does not appear to influence the spin-wave excitations captured in our model.

\subsection{Inelastic Neutron Scattering}\label{Sec:INS}

To understand the magnetic and ferroelectric properties of pyroxenes, particularly the formation of their incommensurate magnetic structures, it is important to determine the underlying magnetic exchange interactions.
INS provides a powerful method for directly probing these interactions by mapping out spin excitations in momentum and energy space.

\begin{figure}[tb]
\center{\includegraphics[width=1.0\linewidth]{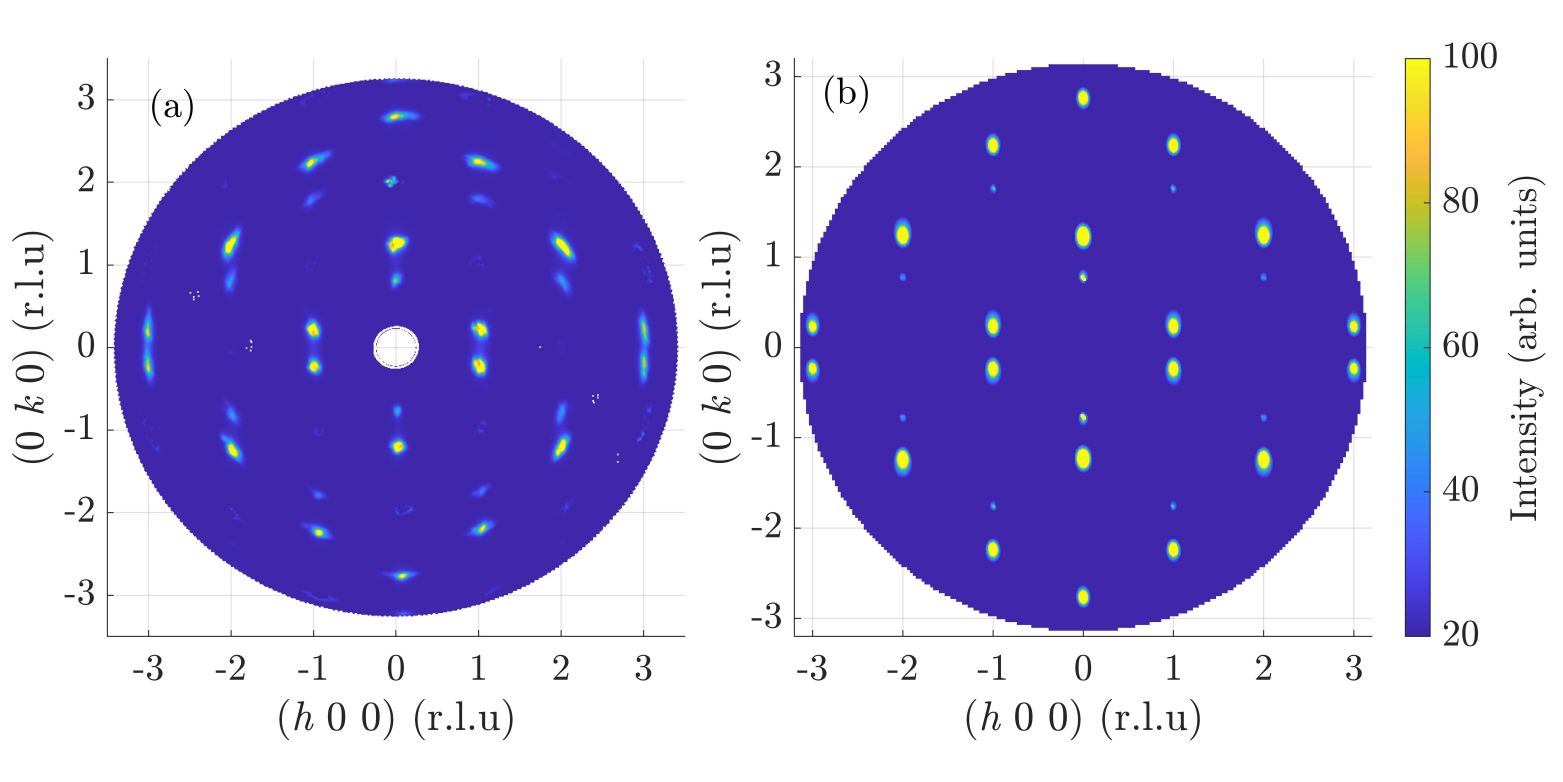}}
  \caption{~(a) The magnetic elastic neutron scattering signal measured in the ($h~k~0$) plane, obtained by subtracting the data at $T=20$~K from the data at $T=1.7$~K, integrated over an energy-transfer range of $E_f = -0.1$ to 0.1~meV. (b) The elastic neutron scattering intensity calculated by LSWT using the best fit parameters of the spin Hamiltonian (Eq.~\ref{XXZ}), see text.
  }
  \label{fig_elastic_tof}
\end{figure}

The TOF INS data measured using the CNCS spectrometer at SNS were recorded in two distinct scattering planes, ($h~k~0$) and ($0~k~l$). We start with the elastic magnetic scattering map obtained by data integration over the energy transfer range -0.1..0.1 meV. Figure~\ref{fig_elastic_tof}(a) shows the corresponding map in the ($h~k~0$) plane, obtained by subtracting high-temperature (paramagnetic) background data at 20~K from data taken in the ordered phase at 1.7~K. 
Similarly to the diffraction measurements above, the ICM magnetic Bragg peaks were found at ($h~k\!\pm\!k_{\rm ICM}~0$) and ($0~k\!\pm\!k_{\rm ICM}~l$) positions around the allowed nuclear peaks, while no incommensurability was detected along the $h$- and $l$-directions.

More importantly, using the TOF technique, enabling us to cover a wider range of reciprocal space, we observe a distinctive pattern in the magnetic peak intensities. 
Although the magnetic satellites near $k = 0$ have roughly equal intensities, those around higher $|k|$ values exhibit a marked asymmetry: the $|k| - k_{\rm ICM}$ peaks are consistently more intense than the corresponding $|k| + k_{\rm ICM}$ ones [see also Fig.~\ref{fig_elastic}(a) and (c)].
Such an asymmetric intensity distribution is related to the specific arrangement of magnetic moments in the structure, e.g. helical order.

\begin{figure*}[tb]
\center{\includegraphics[width=1.0\linewidth]{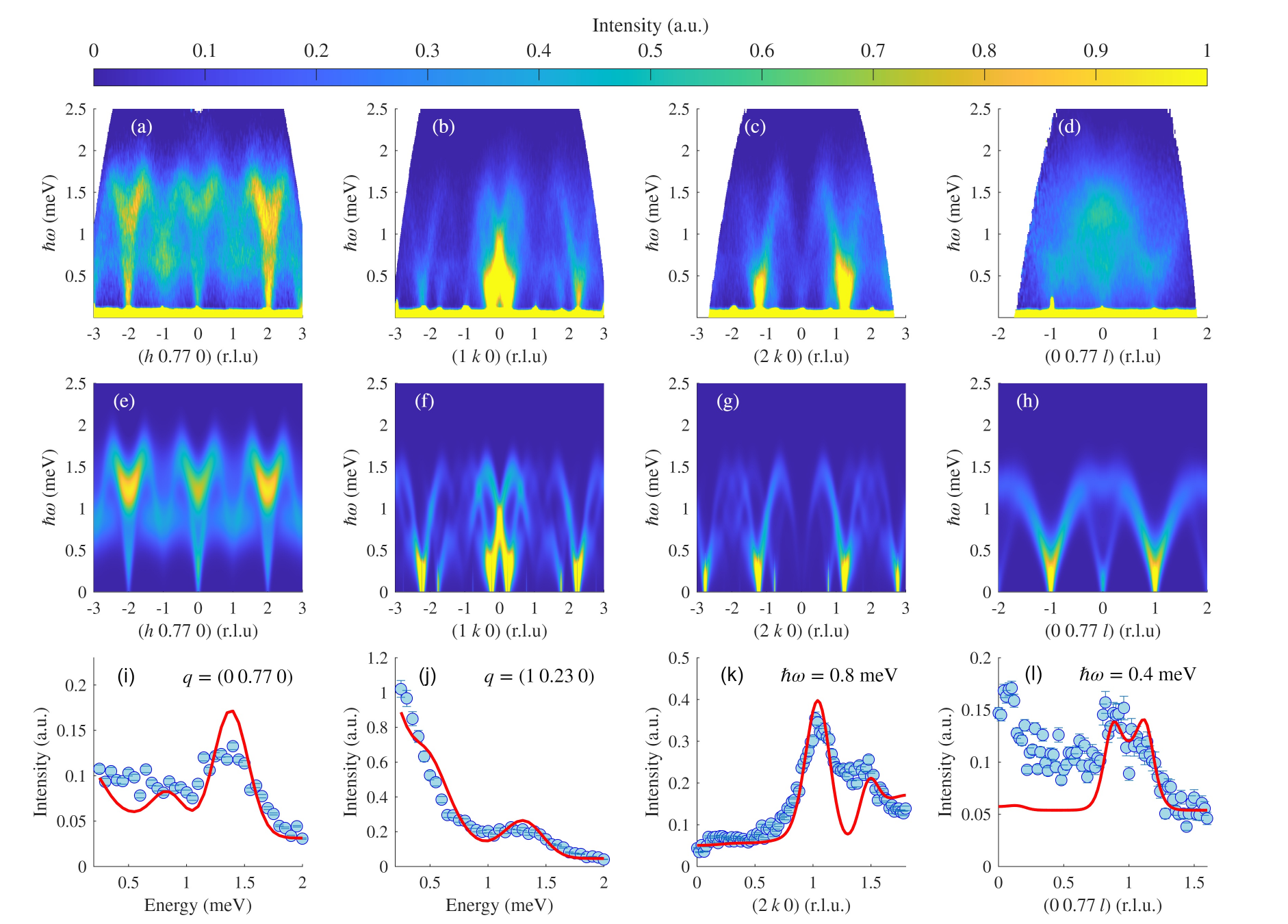}}
  \caption{~Comparison of the experimental spin-wave dispersion and the LSWT calculation.  (a)–(d) INS data measured at $T\!=\!1.7$~K along high symmetry directions in the $(h~k~0)$ and $(0~k~l)$ planes. The data are integrated by $\pm 0.1$~r.l.u. in two orthogonal directions. (e)–(h) The INS intensity (the magnon spectral weight) calculated by LSWT using the best fit parameters of the spin Hamiltonian (Eq.~\ref{XXZ}), see text. (i)–(j) Constant-$\mathbf{Q}$ energy cuts at selected reciprocal space points $(0~0.77~0)$ and $(1~0.23~0)$, respectively.
(k)–(l) Constant-energy cuts at fixed $\hbar\omega = 0.8$~meV and 0.4~meV along $(2~k~0)$ and $(0~0.77~l)$ directions, respectively.
Blue circles show the experimental data, while red lines represent the calculated intensity rescaled by a global fitting factor and offset by a constant background.
}
  \label{ins}
\end{figure*}

Switching to the INS data, we continue with TOF INS measurements. 
Representative experimental momentum- and energy-resolved excitation spectra at $T=1.7$~K are shown in Fig.~\ref{ins}(a-d).
Along the $k$-direction, well-defined spin waves emerge from the ICM Bragg peaks and propagate throughout the Brillouin zone (BZ).
These excitations are gapless and extend up to approximately 1.5 meV.
Additional modes soften at integer $k$ positions, though their intensities are significantly suppressed.

We found that the strong magnon dispersion occurs in all three directions, suggesting significant coupling between spins along all crystallographic axes and a three-dimensional character of the exchange interactions.
Although the broadening is strongest along the $l$-direction, it is also present to some extent along other directions, as seen in Fig.~\ref{ins}. 
This behavior may be related to finite magnon lifetimes or additional damping mechanisms not captured by LSWT.

\subsection{LSWT calculations and discussion}\label{Sec:LSWT}

To validate the experimental findings and determine the exchange parameters, we performed LSWT calculations.
These are shown in Figs.~\ref{ins}(e–h). 
The spin Hamiltonian for a triangular-type magnetic lattice appropriate for \NFS\ and other pyroxenes can be written considering nearest neighbor isotropic intrachain coupling $J$, two interchain couplings $J_1$, $J_2$ (see Fig.~\ref{fig_structure}) and an easy-plane anisotropy term~\cite{Jodlauk_2007,Colin,Biesenkamp2021,Baum2015},
\begin{eqnarray}
  \mathcal{H} &=& J\sum_{ij} \mathbf{S}_i \mathbf{S}_j + J_1\sum_{ij} \mathbf{S}_i \mathbf{S}_j \nonumber \\
  &+&  J_2\sum_{ij} \mathbf{S}_i \mathbf{S}_j + D\sum_i (S^z_i)^2.
  \label{XXZ}
  \end{eqnarray}

The first study of the magnetic structure of aegirine claimed that the intrachain exchange interaction, $J$, is ferromagnetic (FM), while the interchain interactions, $J_1$ and $J_2$, are antiferromagnetic (AFM)~\cite{Ballet1989}.  
Later, local spin-density approximation (LSDA+$U$) calculations~\cite{Streltsov2008} suggested that the three exchange parameters in aegirine are AFM, with the dominant interaction being the intrachain exchange parameter, $J$.
This strong intrachain coupling results in a quasi-one-dimensional (quasi-1D) magnetic system, where the spin dynamics is primarily governed by interactions along the chains, while weaker interchain couplings contribute to the overall three-dimensional magnetic behavior.
However, the authors noted that because of the large oxygen orbitals' polarization, the exchange parameters can be significantly overestimated.

In a system with long-range magnetic order and a large spin $S=5/2$, low-energy excitations can be well described as conventional magnons, representing coherent precessions of spins around the ordered ground state.
LSWT can qualitatively capture the dispersion and magnon bandwidth of ordered magnets throughout the BZ.
In addition to being computationally efficient, LSWT calculations are useful for testing different scenarios, including systems with multiple exchange interactions.

To cover all relevant parameter space between the intrachain and interchain interactions, we allowed the exchange parameters $J$, $J_1$ and $J_2$ to take both FM and AFM values.
We used experimental INS spectra $\hbar\omega < 2$~meV, which exhibit a number of key features, such as the stiffness of the magnon branches, the ICM wave vector, and the magnon spectral weight, to refine the parameters of the Hamiltonian~(\ref{XXZ}).
We quantified the energy of the well-defined magnon modes along different high-symmetry directions in reciprocal space and used this dataset to fit the exchange parameters.
The chi-square fit of the experimentally observed and calculated INS spectra yielded the exchange parameters $J\!=\!0.142(2)$~meV, $J_1\!=\!0.083(1)$~meV, $J_2\!=\!0.186(1)$~meV.
The easy-plane anisotropy $D$ was included as a free parameter in the fit and converged to a small but finite value, $D=0.020(6)$~meV, which ensures that the magnetic moments remain confined within the $ac$ plane.
The calculated spectra show good agreement with the experimental data [Fig.~\ref{ins}(a)-(h)], successfully capturing both the dispersion relations and the spectral weight distribution across reciprocal space.

To further evaluate the accuracy of the LSWT model, we performed a quantitative comparison between calculated and experimental constant-Q and constant-energy cuts at representative positions in reciprocal space, shown in Figs.~\ref{ins}(i)–(l). 
The theoretical intensity was scaled using a global fitting factor, with a constant background added. 
The calculated peaks remain systematically narrower than the experimental ones, especially for the excitations along the $l$ direction. 
This broadening may be attributed to magnon–magnon interactions or finite magnon lifetimes not reproduced by LSWT. 
Despite this, the observed agreement in both the energy and Q-cuts supports the validity of the extracted exchange parameters in describing the spin dynamics of \NFS.

The fitted interchain exchange parameters $J_1$ and $J_2$  correspond to Fe–Fe couplings mediated by one and two SiO$_2$ tetrahedra, respectively. 
$J_2$ is found to be roughly twice as large as $J_1$, although the Fe–Fe distance is longer. 
Similar behavior was reported for related pyroxenes such as LiFeSi$_2$O$_6$, and NaCrGe$_2$O$_6$ where it was explained in terms of the exchange pathway geometry~\cite{Jodlauk_2007,Streltsov2008}. 
In particular, the $J_2$ path, which runs through two edge-sharing tetrahedra, provides more favorable orbital overlap and symmetry conditions that enhance the AFM superexchange. 
In contrast, the $J_1$ path involves a single tetrahedron with more distorted geometry and less effective overlap. 
All three exchange constants are antiferromagnetic and consistent with the Goodenough–Kanamori rules for superexchange between half-filled $d^5$  ions. The Fe–Fe distances associated with each exchange path are $J$: 3.19~\AA, $J_1$: 6.53~\AA, and $J_2$: 5.44~\AA.

Although the exchange couplings $J$, $J_1$, and $J_2$ differ in strength, their AFM character and comparable magnitudes give rise to competing interactions.
This frustration destabilizes simple collinear spin arrangements and favors the formation of an incommensurate helical magnetic structure.

\section{Conclusion}

Our study of spin dynamics in the natural mineral aegirine \NFS\ provides a detailed and quantitative understanding of low-energy magnetic excitations in the helical phase of this multiferroic material. 
Using LSWT, we successfully reproduced the observed INS spectra and extracted a consistent set of key exchange interactions that describes the helical ground state and magnetic excitations.

We found that all exchange interactions are AFM, in agreement with the LSDA calculations~\cite{Streltsov2008} and the competition between of them stabilizes the helical ground state.
This competition may be responsible for the reduction of the ordering temperature and a moderate frustration parameter, $f = 4.3$ observed in magnetization measurements~\cite{baum1988structural}. 
Surprisingly, the strongest coupling, $J_2$, connects spins between the chains, whereas the intrachain exchange, $J$, is approximately 25\% weaker. 
Our findings revise the previously suggested quasi-one-dimensional picture of aegirine. The dominant interchain exchange $J_2$, along with two other comparable antiferromagnetic couplings, reveals that spin interactions are inherently three-dimensional. 

This highlights the need to go beyond simple structural intuition and apply detailed spectroscopic analysis when characterizing frustrated magnetic systems.

Future INS experiments under applied magnetic fields will be essential for mapping the full magnetic phase diagram and refining the spin Hamiltonian, particularly with respect to anisotropic interactions.

\section*{Acknowledgments}

O.P. acknowledges the GIMRT Program of the Institute for Materials Research, Tohoku University (Proposals No. 202305-HMKPB-0517, No. 202403-HMKPB-0513, No. 202305-CNKXX-0508, and No. 202403-CNKXX-0503). 
Work by L.M.A. was supported by the U.S. Department of Energy, Office of Science, Office of Basic Energy Sciences, Chemical Sciences, Geosciences, \& Biosciences (CSGB) Division.
A part of this work was supported by JSPS KAKENHI JP21H04987(K.K. and M.F.), JP23H04867(C.T.), and JP23H04871(Y.H.).
This research used resources at the Spallation Neutron Source, a DOE Office of Science User Facility operated by Oak Ridge National Laboratory. The beam time was allocated to CNCS on proposal number IPTS-33211. 
The experiment on TAS-2 was performed under the US-Japan Cooperative Program on Neutron Scattering and Proposal No. D1101.
This work is partially based on experiments performed at the Swiss spallation neutron source SINQ, Paul Scherrer Institute, Villigen, Switzerland. 

\section*{Data availability}

INS data supporting the findings of this article are openly available~\cite{Datatof,dataJRR-3}. 
The rest of the data are available upon reasonable request from the authors.

\bibliography{bibliography}

\end{document}